\documentclass[preprint,12pt]{elsarticle}

\usepackage{amssymb}
\usepackage{hyperref}

\newcounter{bla}

\usepackage{subfig}
\usepackage{amssymb,color}

\newcommand{\blue}{\textcolor{blue}}
\journal{Computer Physics and Communications}

\begin{document}

\begin{frontmatter}




\title{GDoeSII: Open source software for design of diffractive optical elements and conversion to GDSII lithography files}


\author[a,b]{Raghu Dharmavarapu\corref{author}}
\author[a]{Shanti Bhattacharya}
\author[b]{Saulius Juodkazis}

\cortext[author] {Corresponding author.\\\textit{E-mail address:} raghu.d@ee.iitm.ac.in}
\address[a]{Centre for NEMS and Nanophotonics (CNNP), Department of Electrical Engineering, Indian Institute of Technology Madras, Chennai 600036, India}
\address[b]{Centre for Micro-Photonics, Faculty of Science, Engineering and Technology, Swinburne University of Technology, Hawthorn VIC 3122, Australia}

\begin{abstract}
We develop an open source software ``\emph{GDoeSII}'' for the simulation of Fresnel (near field) and Fraunhofer (far field) diffraction integrals of diffractive optical elements (DOE). It can compute the intensity  distributions at a desired plane from the DOE. This software can also convert the phase profiles of the DOEs from standard image formats such as JPG and PNG to lithography graphic format GDSII for fabrication purposes. The conversion algorithm used in this program groups adjacent same-valued pixels and creates a single cell rather than creating a cell for each and every pixel in the image file. This results in a faster and much lower output GDSII file size. Conversion to multi-layer GDSII format is also possible for grayscale lithography. We show as an example: The simulation, GDSII conversion, fabrication and experimental results for the case of an Airy beam generator. This application offers a complete platform for researchers working in diffractive optics from simulations to lithography file preparation.
\end{abstract}


\end{frontmatter}


{\bf PROGRAM SUMMARY/NEW VERSION PROGRAM SUMMARY}

\begin{small}
\noindent
{\em Program Title:} GDoeSII             \\
\blue{Download software here:} \blue{\href{https://drive.google.com/file/d/1srE3iXJumPXlDW0RtBuIYfG7zvtYAFhm/view?usp=sharing}{\underline Download} }
\\
{\em Licensing provisions(please choose one):} CC by 4.0                \\
{\em Programming language:} Python                                  \\ 

{\em External routines/libraries:} gdsCAD, NumPy, SciPy, Tkinter \\

{\em Supplementary material:}                                 \\
{\em Journal reference of previous version:}                  \\
{\em Does the new version supersede the previous version?:}   \\
{\em Reasons for the new version:}\\
{\em Summary of revisions:}*\\

{\em Nature of problem(approx. 50-250 words):}\\
Computing the complex light field in a plane at a distance from Diffractive optical elements (DOE) and converting the DOE phase profile images into GDSII format for lithography.\\

{\em Solution method(approx. 50-250 words):}\\
We use Fresnel and Fraunhofer scalar diffraction integrals to compute the complex electric field of light at a distance from the plane of the DOE. We designed an algorithm to convert the DOE phase images into GDSII format with low output file size and shorter conversion times using Python.\\

{\em Additional comments including Restrictions and Unusual features (approx. 50-250 words):}\\
   \\

* Items marked with an asterisk are only required for new versions
of programs previously published in the CPC Program Library.\\
\end{small}

\section{Introduction}
Diffractive optical elements (DOE) are thin phase or amplitude elements that operate by means of diffraction to produce arbitrary distributions of light. Some examples of DOEs include diffraction gratings, Fresnel zone lenses~\cite{shiono1990blazed}, diffractive axicons~\cite{golub2006fresnel} and so on. DOEs are extensively used as beam shaping elements, lenses and other essential optical components. The typical design flow of DOEs is as follows: 1. Calculation of DOE phase profile, using techniques such as the G-S algorithm~\cite{gerchberg1972practical}, simplified mesh technique~\cite{bhattacharya2008simplified}, modulo $2\pi$ conversion of refractive elements or analytic equations~\cite{bhattacharya2017design} 2. Simulating the optical fields using scalar diffraction equations and 3. Fabrication using lithography techniques followed by experimental verification.

In this paper we present the implementation details of \textit{GDoeSII}, an all-in-one software with graphical user interface, which allows users to perform scalar diffraction simulations and then convert the phase profiles from standard image formats such as JPG and PNG to GDSII format for further lithography process. We believe this software will help graduate students and researchers working in the field of difractive optics and allied fields. The following sections explain the implementation details of the two main modules of the software which are: 1. DOE module, 2. GDSII conversion module.
\section{Diffractive optics module}
\begin{figure}[!h]
\centering
\includegraphics[width = 11cm]{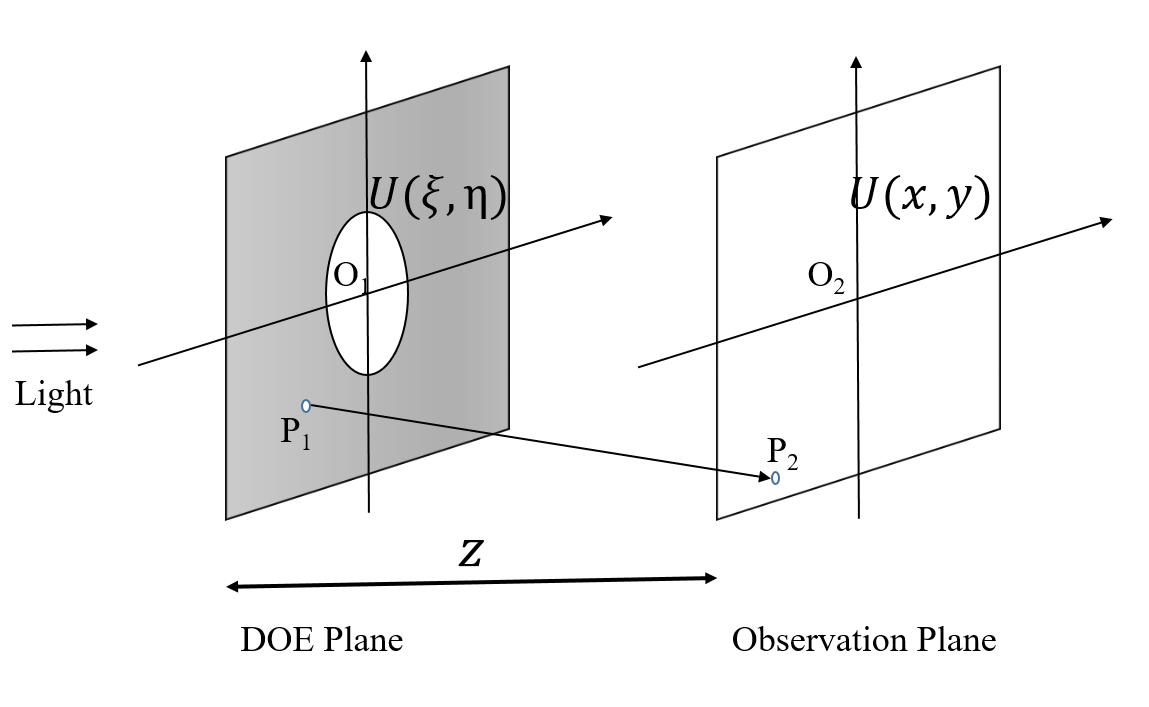}
\caption{Geometry of the planes for scalar diffraction theory}
\label{fig:1}
\end{figure}

In this section we discuss the theoretical background behind the DOE simulator module. The geometry of the problem is shown in Fig.~\ref{fig:1}. We used the scalar diffraction integrals which are  the Fresnel and Fraunhofer diffraction integrals to compute the intensity of the field distribution at a desired plane from the DOE. The software uses the Fresnel (\textit{near field}) diffraction integral Eq.~\ref{eq:1} to compute the complex field $U(x,y,z)$ at a distance $z$ from the DOE plane.
\begin{equation}
U(x,y,z) = \frac{e^{ikz}}{j\lambda z}e^{j\frac{k}{2z}(x^2+y^2)}\int \int \{ U(\xi , \eta) e^{j\frac{k}{2z}(\xi^2+\eta^2)}\}e^{-j\frac{2\pi}{\lambda z}(x\xi + y\eta)}d\xi d\eta,
\label{eq:1}
\end{equation}
Where $x$, $y$ are the coordinate system in the observation plane and $\xi,\eta$ are the coordinate system in the DOE plane,  $k$ is the wave vector and $\lambda$ is the wavelength of the light.
A simplified version of Eq.~\ref{eq:1}, also known as Fraunhofer diffraction(\textit{far field}) integral given in Eq.~\ref{eq:2} is used to compute the complex field at far field.
\begin{equation}
U(x,y) = \frac{e^{jkz}e^{j\frac{k}{2z}(x^2+y^2)}}{j\lambda z}\int \int U(\xi,\eta)e^{-j\frac{2\pi}{\lambda z}(x\xi+y\eta)}d\xi d\eta
\label{eq:2}
\end{equation}
The above two equations yield reasonably accurate result under the following conditions~\cite{goodman2005introduction}:
\begin{enumerate}
\item The diffracting aperture must be large compared with the wavelength $\lambda$
\item The diffracting fields must not be observed too close to the aperture
\end{enumerate}

\section{GDSII conversion module}
The next step after simulation is the fabrication of the physical DOE. Most lithography systems such as the Electron beam lithography (EBL) and Photo/UV lithography systems accept the DOE phase designs in only GDSII format. Often researchers have to rely on expensive software for this conversion step. 
\par
\begin{figure}[!h]
\centering
\includegraphics[width = 2in]{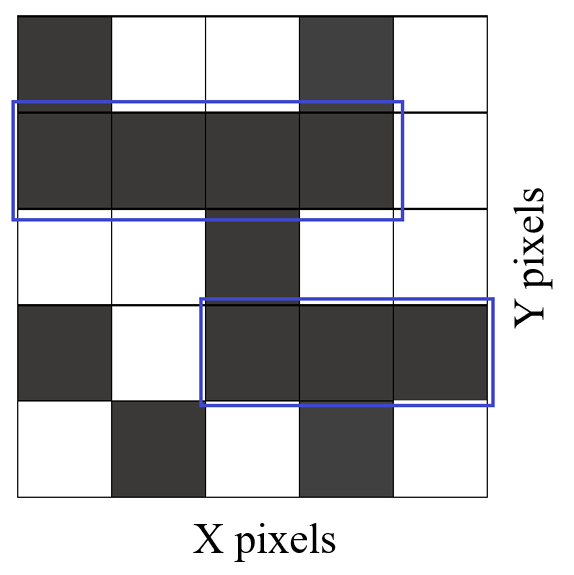}
\caption{Grouping of same valued pixels in a row}
\label{fig:grouping}
\end{figure}
\begin{figure}[!h]
\centering
\subfloat[]{\includegraphics[height=1.75in]{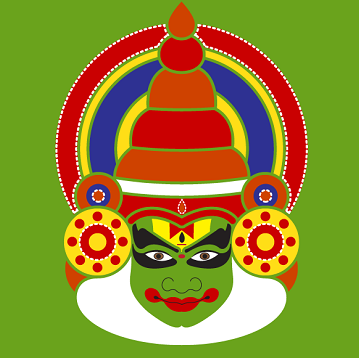}}
\hspace{0.5cm}
\subfloat[]{\includegraphics[height=1.75in]{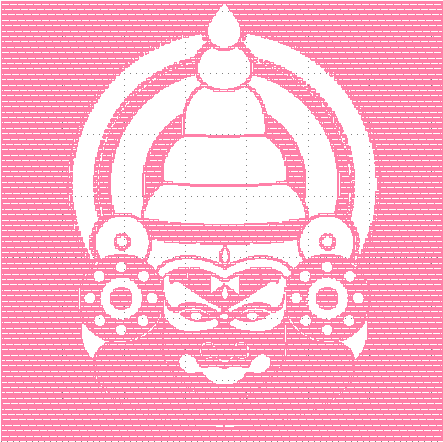}}
\\
\subfloat[]{\includegraphics[height=1.75in]{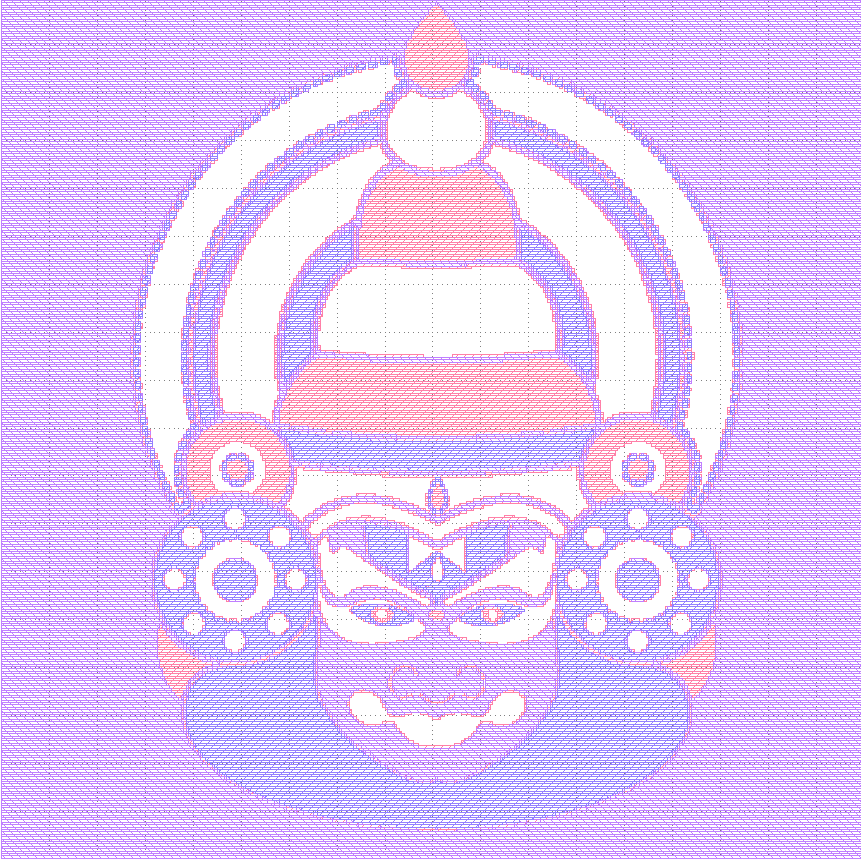}}
\hspace{0.5cm}
\subfloat[]{\includegraphics[height=1.75in]{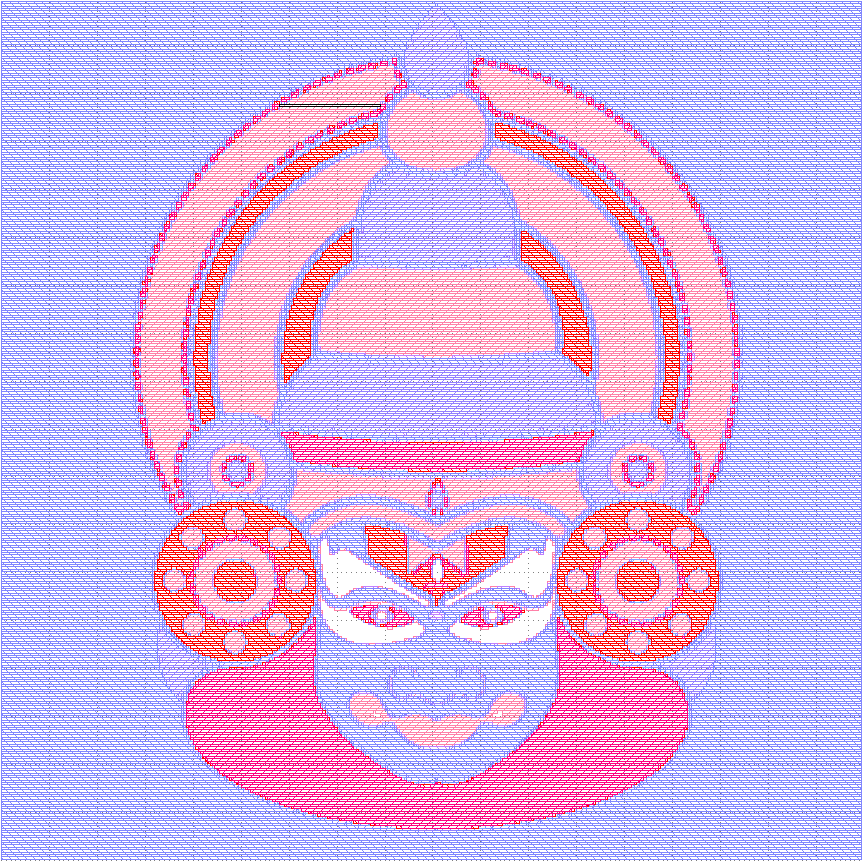}}
\caption{(a) Original image, Converted GDSII files (b) 2 level (c) 4 level (d) 8 levels}
\label{fig:2}
\end{figure}

To address this important step in the DOE design cycle, we added the GDSII conversion module for the conversion of the DOE phase profile from standard image formats such as JPG and PNG to GDSII format. Typical image to GDSII conversion process involves creating a square pixel in the GDS file for each white or dark colored pixel in the image file. This process takes very long time and the output GDS file size is usually very large. In our approach, the algorithm detects continuous same valued pixels and groups them as a line segment in the GDS file. The line segment takes much less memory. The grouping method used in our program is depicted in Fig.~\ref{fig:grouping}. The algorithm can also convert the image file into $n$ layered GDSII files by quantizing the image to $n$ intensity levels. This is very useful for researchers working with 3D optical elements. Fig.\ref{fig:2} shows the images converted using \textit{GDoeSII} into 2, 4, 8 levels respectively. This module offers another feature to create arrays of basic shapes such as circles, triangles and rectangles.
\par
\begin{figure}
\centering
\subfloat[]{\includegraphics[width  = 7 cm]{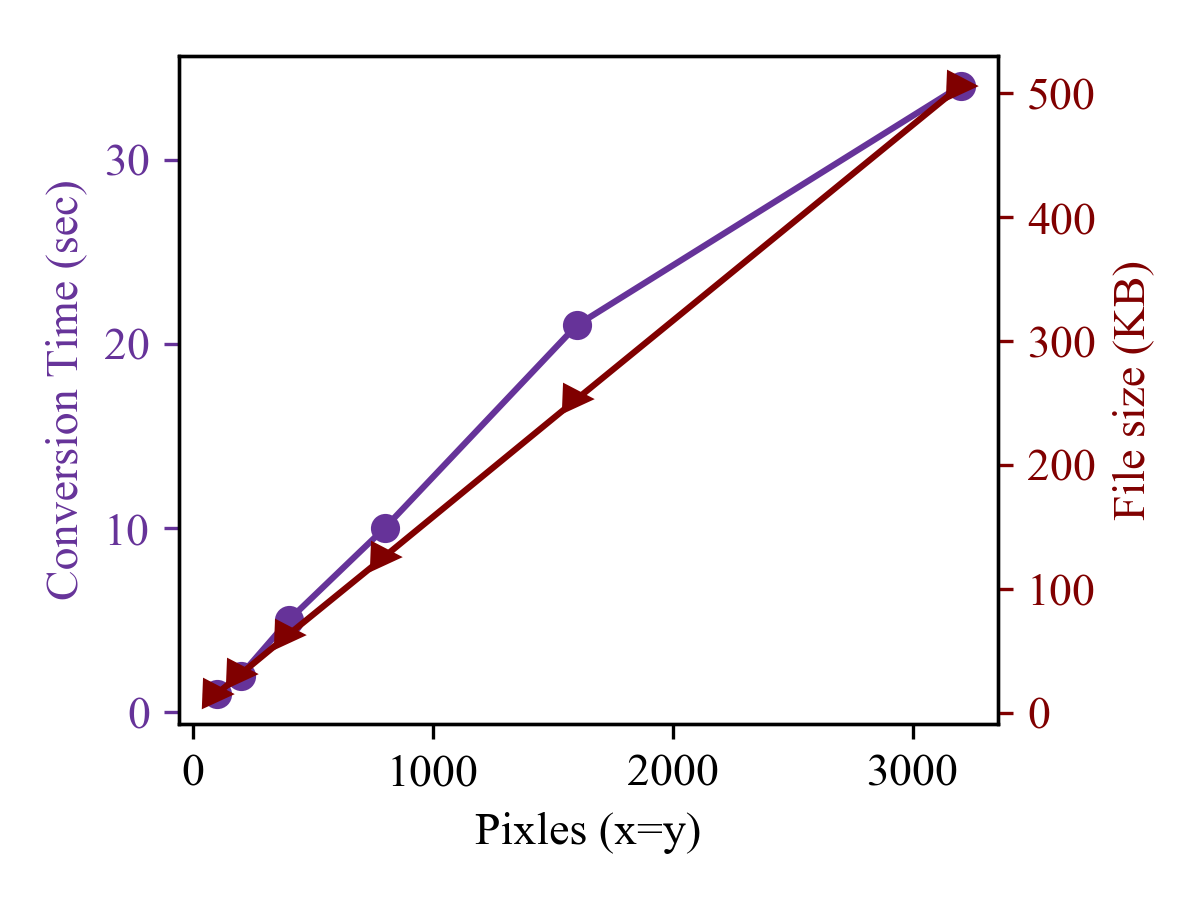}}
\subfloat[]{\includegraphics[width  = 7 cm]{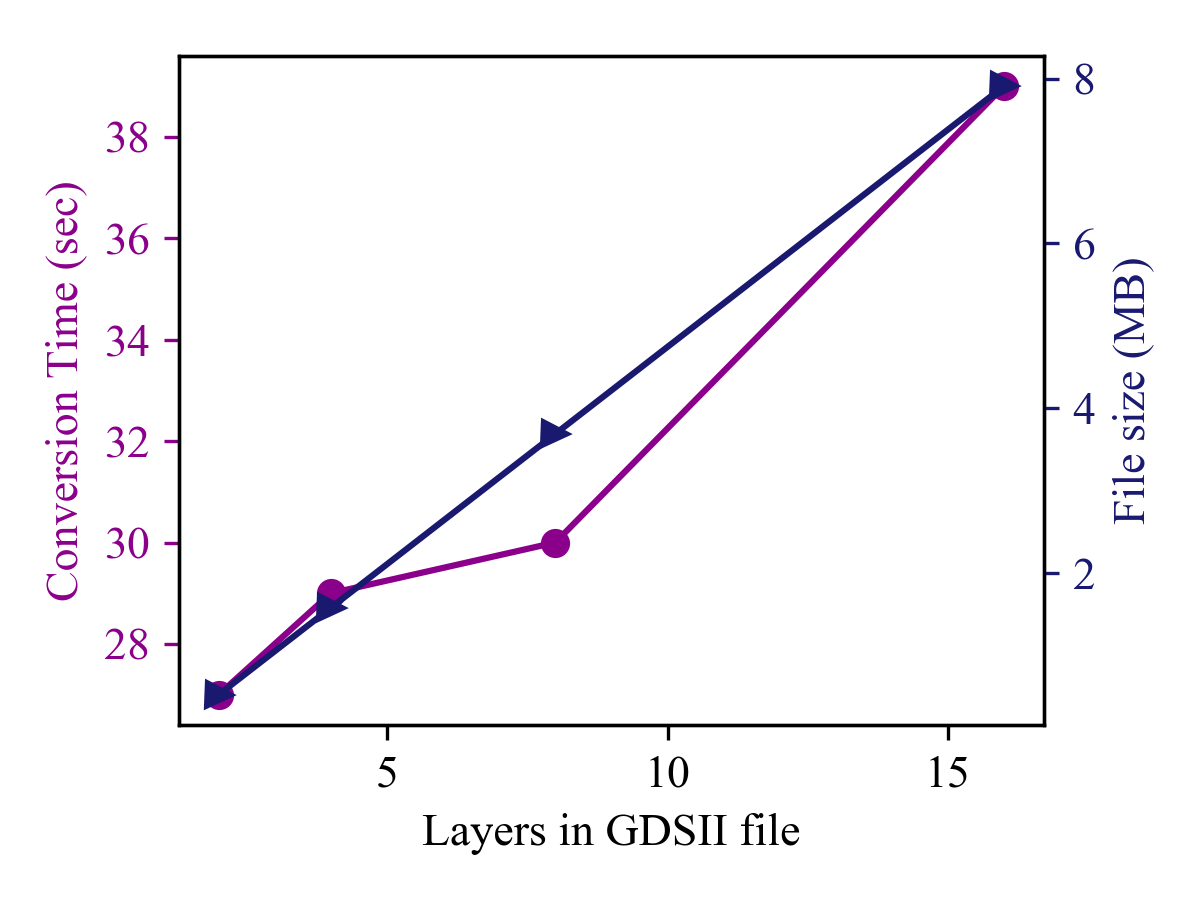}}
\caption{GDS file conversion time and output file size(a) for different image sizes  (b) for same image but different number of layers in output GDS file}
\label{fig:metrics}
\end{figure}

The following parameters were measured to access the performance of the software. Fig.~\ref{fig:metrics} (a) shows the conversion times and output GDS file size for different sizes of the input image. Fig.~\ref{fig:metrics} (b) shows the conversion time and output GDS file size when a single image (2000x2000) was converted to multiple layered GDS file. However, these numbers can vary with the hardware capacity of the machine.

\section{Simulation and Experiment results}
\begin{figure}[!h]
\centering
\subfloat[]{\includegraphics[height = 1.5in]{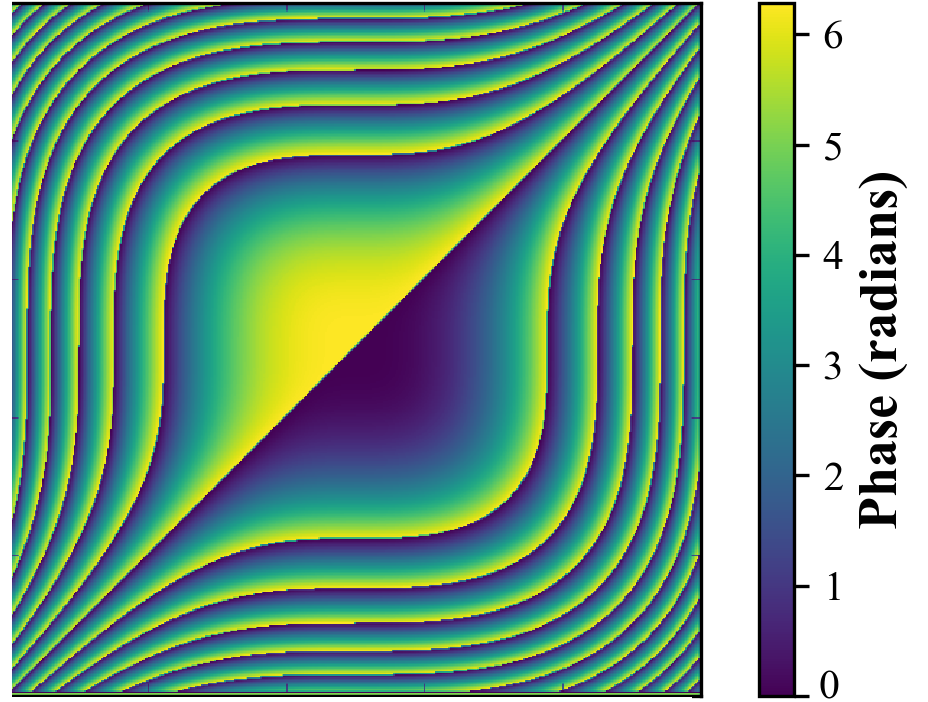}}
\hspace{0.5cm}
\subfloat[]{\includegraphics[height = 1.5in]{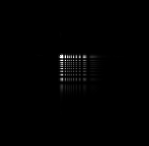}}
\caption{(a) Cubic phase profile (b) Airy beam Intensity at $z$ = 10~cm (Fresnel)}
\label{fig:3}
\end{figure}

In this section we show the simulation, fabrication and experimental results for the case of an Airy beam~\cite{siviloglou2007observation}. An Airy beam is a non diffractive solution of the paraxial diffraction equation. The airy beam can be generated using a cubic phase profile, which is shown in Fig. \ref{fig:3}.
\par
\begin{figure}[!h]
\centering
\subfloat[]{\includegraphics[height = 1.5in]{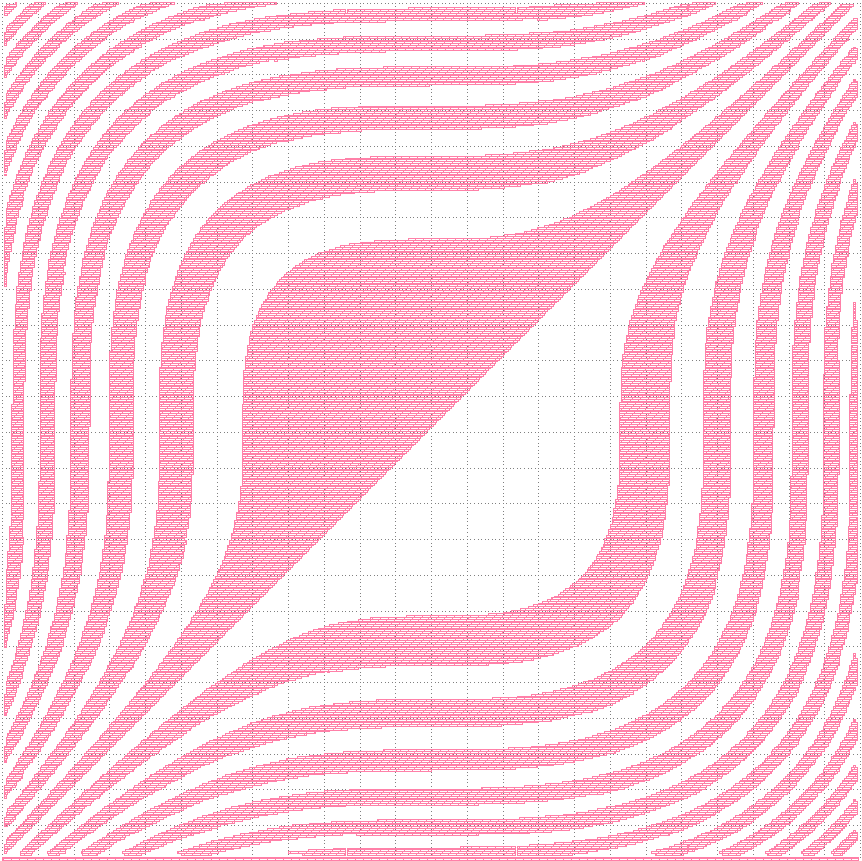}}
\hspace{0.1cm}
\subfloat[]{\includegraphics[height = 1.5in]{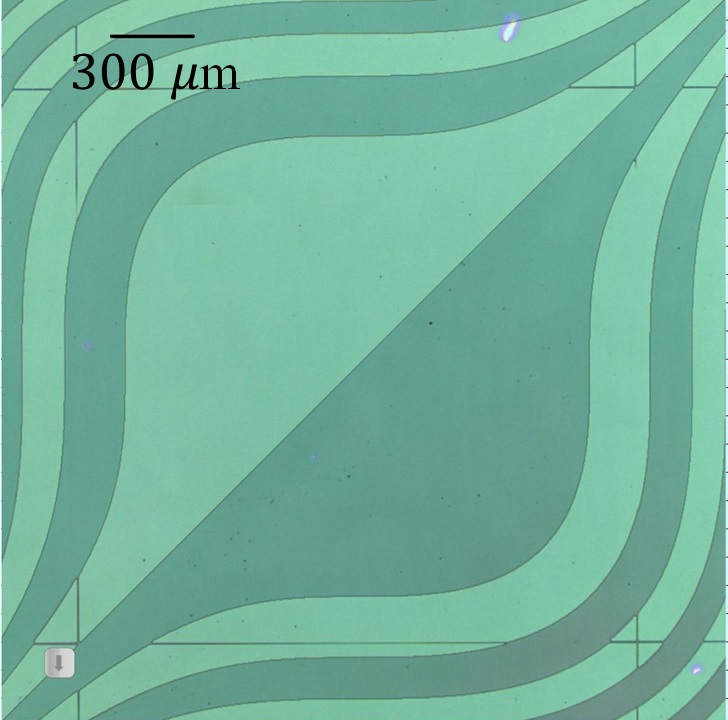}}
\hspace{0.1cm}
\subfloat[]{\includegraphics[height = 1.5in]{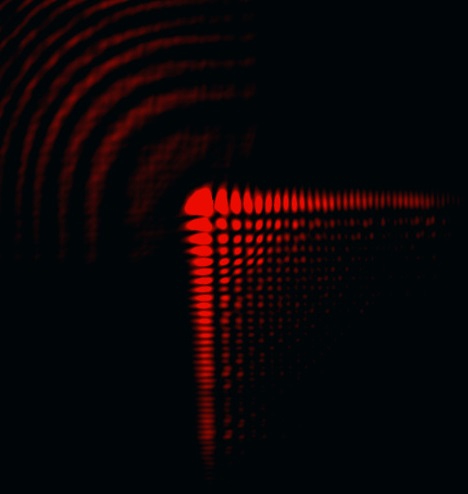}}
\caption{(a) GDSII converted image (b) Confocal microscope image of the DOE (c) Experimental CCD image of the Airy beam}
\label{fig:4}
\end{figure}

The phase profile shown in Fig.~\ref{fig:3} is converted into GDSII format and fabricated using Electron beam lithography (Raith 150 TWO) system. These results are summarized in Fig.~\ref{fig:4}. Fig.~\ref{fig:4} (c) shows the experimentally generated Airy beam which matches the simulated intensity profile shown in Fig.~\ref{fig:3} (b).
\begin{figure}[!h]
\centering\includegraphics[height = 1.5in]{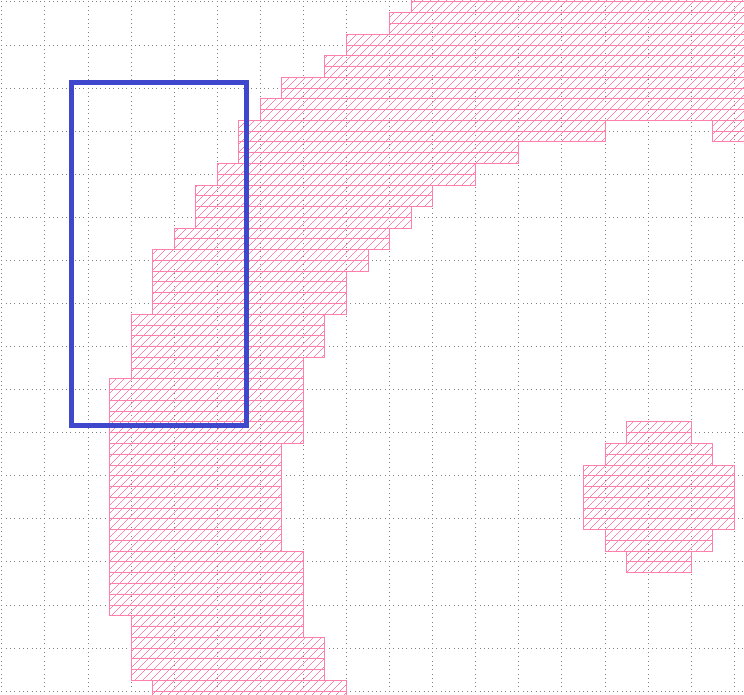}
\caption{Curved region in an example GDSII design}
\label{fig:5}
\end{figure}
\section{Conclusion}
We have introduced \textit{GDoeSII}, a Python based software for Microsoft Windows platform which facilitates the computation of Intensity distributions produced by diffractive optical elements. This program also enables the users to convert phase profiles from image formats such as JPG and PNG to GDSII (also multi layer) format in very less time and file size. One problem with this conversion arises when there are curved features in the image file. Fig.~\ref{fig:5} shows a curved segment of the GDSII file which clearly shows rough ridges. Future updates to the software will focus on further improvement of the algorithm to accurately convert curved elements.

We hope that our openly available software will help researchers in the field of Optics and Nanofabrication. 
\section{Acknowledgements}
\noindent
R D thanks Dr. Vijaya Kumar, Dr. Soon Hock Ng, Vandhana Narayanan and Sripriya for their contribution in testing the software.
\label{}
\\



\section*{References}
\bibliographystyle{elsarticle-num}
\bibliography{bib}

\begin{thebibliography}{1}
\expandafter\ifx\csname url\endcsname\relax
  \def\url#1{\texttt{#1}}\fi
\expandafter\ifx\csname urlprefix\endcsname\relax\def\urlprefix{URL }\fi
\expandafter\ifx\csname href\endcsname\relax
  \def\href#1#2{#2} \def\path#1{#1}\fi

\bibitem{shiono1990blazed}
T.~Shiono, K.~Setsune, Blazed reflection micro-fresnel lenses fabricated by
  electron-beam writing and dry development, Optics letters 15~(1) (1990)
  84--86.

\bibitem{golub2006fresnel}
I.~Golub, Fresnel axicon, Optics letters 31~(12) (2006) 1890--1892.

\bibitem{gerchberg1972practical}
R.~Gerchberg, W.~Saxton, A practical algorithm for the determination of the
  phase from image and diffraction plane pictures, Optik (Jena) 35 (1972) 237.

\bibitem{bhattacharya2008simplified}
S.~Bhattacharya, Simplified mesh techniques for design of beam-shaping
  diffractive optical elements, Optik-International Journal for Light and
  Electron Optics 119~(7) (2008) 321--328.

\bibitem{bhattacharya2017design}
S.~Bhattacharya, A.~Vijayakumar,
  \href{https://books.google.com.au/books?id=tGcEMQAACAAJ}{Design and
  Fabrication of Diffractive Optical Elements with MATLAB}, Tutorial Texts,
  SPIE Press, 2017.
\newline\urlprefix\url{https://books.google.com.au/books?id=tGcEMQAACAAJ}

\bibitem{goodman2005introduction}
J.~W. Goodman, Chapter 4, Introduction to Fourier optics, Roberts and Company
  Publishers, 2005.

\bibitem{siviloglou2007observation}
G.~Siviloglou, J.~Broky, A.~Dogariu, D.~Christodoulides, Observation of
  accelerating airy beams, Physical Review Letters 99~(21) (2007) 213901.

\end{thebibliography}







\end{document}